\documentclass[preprint,authoryear,12pt]{elsarticle}

\usepackage{lineno,hyperref}
\modulolinenumbers[5]

\begin{document}
\begin{frontmatter}

\title{The treatment of magnetic buoyancy in flux transport dynamo models}
\author[mymainaddress]{Arnab Rai Choudhuri\corref{mycorrespondingauthor}}
\cortext[mycorrespondingauthor]{Corresponding author}
\ead{arnab@physics.iisc.ernet.in}

\author[mymainaddress,mysecondaryaddress]{Gopal Hazra}
\ead{ghazra@physics.iisc.ernet.in}

\address[mymainaddress]{Department of Physics, Indian Institute of Science,Bangalore 560012, India}
\address[mysecondaryaddress]{Indian Institute of Astrophysics, Bangalore-560034, India}

\begin{abstract}

One important ingredient of flux transport dynamo models is
the rise of the toroidal magnetic field through the convection
zone due to magnetic buoyancy to produce bipolar sunspots and
then the generation of the poloidal magnetic field from these
bipolar sunspots due to the Babcock--Leighton mechanism.
Over the years, two methods of treating magnetic buoyancy---a
local method and a non-local method---have been used widely
by different groups in constructing 2D kinematic models of
the flux transport dynamo.  We review both these methods and
conclude that neither of them is fully satisfactory---presumably
because magnetic buoyancy is an inherently 3D process.  We also
point out so far we do not have proper understanding of why
sunspot emergence is restricted to rather low latitudes.

\end{abstract}
\end{frontmatter}


\section{Introduction}

From 1990s a type of model has been developed for the solar
dynamo known as the flux transport dynamo model
\citep{WSN91,CSD95,Durney95}. At the present time, 
this model seems to be the most promising and satisfactory
model for explaining different aspects of the solar cycle
\citep{Chou11,Chou15, Charbonneau14}, although still doubts are
sometimes expressed about its validity.
The flux transport dynamo model basically involves
the following three processes.
(i) The strong toroidal field is produced by the
stretching of the poloidal field by differential rotation in
the tachocline. (ii) The toroidal field generated in the
tachocline gives rise to active regions due to magnetic
buoyancy, and the decay of tilted bipolar active regions
produces the poloidal field by the Babcock--Leighton
mechanism. (iii) The
poloidal field produced by the Babcock--Leighton mechanism
is advected by the meridional circulation first to high latitudes 
and then down to the tachocline, while also diffusing down to
the tachocline due to turbulent diffusion.

Most of the calculations of the flux transport dynamo model
have been based on axisymmetric 2D kinematic mean field
equations.  It is completely straightforward to include
the process (i) within such a formalism---especially because
helioseismology gives us the profile of differential rotation.
As far as process (iii) is concerned, there are some uncertainties
in the nature of the meridional circulation \citep{Hathaway12,Zhao13,Schad13,RA15} as well as in the
value of turbulent diffusion \citep{Jiang07,Yeates08}.  Although early models
assumed a simple one-cell form of meridional circulation,
there have been some recent calculations with more
complicated meridional circulation \citep{HKC14}. However, once
we specify the form of the meridional circulation and the turbulent
diffusion within the framework of the kinematic model, there
is absolutely no uncertainty in the mathematical forms of
the terms involved in process (iii). Only in the case of
process (ii) involving magnetic buoyancy and the Babcock--Leighton
mechanism, there is considerable
uncertainty at a fundamental level as to how this process should 
be included in a 2D kinematic model.  Magnetic buoyancy involves
the rise of a tilted flux loop through the convection zone and
is an inherently 3D process, which can be included in a 2D dynamo
model only through rather crude approximation procedures.
Over the years, different groups have proposed different
procedures for handling the process (ii). While carrying on
calculations with the flux transport dynamo model, we have
become aware that these different procedures often give significantly
different results and we have tried to understand the physical
reasons behind these differences.  The obvious question is: which
one is the most realistic procedure for treating the process (ii)?
This is not an easy question to settle. Different procedures have
their own strengths and own weaknesses.  We have found that the
subtleties involved in modelling the process (ii) are not
sufficiently appreciated by the scientific community. 
Probably a fully satisfactory treatment of process (ii) is not
possible within the 2D kinematic framework and one has to go
beyond 2D. We review 
the different procedures which had been proposed by different
groups and critically examine the aspects of physics which are
covered and which are not covered in these different procedures.

\def\vb{{\bf v}}
\def\Bb{{\bf B}}
\def\ol{\overline}
\def\ec{\cal{E}}
\def\pa{\partial}
\def\vf{{\bf v}}
\def\Bf{{\bf B}}

\section{The basic equations}

We assume both the mean magnetic field and the mean velocity
field to be axisymmetric in 2D kinematic models. The magnetic
field is written as
$${\bf B} = B (r, \theta) {\bf e}_{\phi} + \nabla \times [ A
(r, \theta) {\bf e}_{\phi}],
\eqno(1)$$
where $B (r, \theta)$ is the toroidal component and $A (r, \theta)$
gives the poloidal component. We can write the velocity field
as $\vb + r \sin \theta \, \Omega (r, \theta) {\bf e}_{
\phi}$, where 
$\Omega (r, \theta)$ is the angular velocity in the interior of the
Sun and $\vb$ is the velocity of meridional circulation having components
in $r$ and $\theta$ directions.  Then the main equations telling us
how the poloidal and the toroidal fields evolve with time are
$$\frac{\pa A}{\pa t} + \frac{1}{s}(\vf.\nabla)(s A)
= \lambda_T \left( \nabla^2 - \frac{1}{s^2} \right) A + S(r, \theta, t),
\eqno(2)$$
$$ \frac{\pa B}{\pa t} 
+ \frac{1}{r} \left[ \frac{\pa}{\pa r}
(r v_r B) + \frac{\pa}{\pa \theta}(v_{\theta} B) \right]
= \lambda_T \left( \nabla^2 - \frac{1}{s^2} \right) B 
+ s(\Bf_p.\nabla)\Omega + \frac{1}{r}\frac{d\lambda_T}{dr}
\frac{\partial}{\partial{r}}(r B), \eqno(3)$$
where $s = r \sin \theta$, $\lambda_T$ is the turbulent
diffusivity and $S(r,\theta, t)$ is the dynamo source term.

The term $s(\Bf_p.\nabla)\Omega$ in (3) corresponds to process
(i) involving the generation of the toroidal field from the
poloidal field involving differential rotation.  On the other
hand, the terms $s^{-1}(\vf.\nabla)(s A)$ and
$\lambda_T \left( \nabla^2 - 1/ s^2 \right) A$ in (2) correspond
to process (iii) involving the evolution of the poloidal field
due to the meridional circulation and the turbulent diffusion
together. It is the source term $S(r,\theta, t)$ which incorporates
the process (ii).  Sometimes we have to do some extra things
to (2) as well (as described below) in order to include the
magnetic buoyancy of the toroidal field $B$.

In the early $\alpha \Omega$ dynamo model postulated by
\citet{Parker55a} and \citet{SKR66}, the
source term is $S = \alpha B$ (see, for example, \citet{Chou98}, Chapter 16). Here $\alpha$ is a measure of helical turbulence and is usually referred as the $\alpha$-effect.
Although the Babcock--Leighton mechanism also can be encaptured
by a superficially similar $\alpha$-coefficient, its physical
interpretation is completely different from that of the
$\alpha$-effect, which implies the twisting of the toroidal
field. When it was realized that the toroidal field at the
bottom of the convection zone is much stronger than what was
assumed earlier \citep{CG87,chou89,Dsilva93,Fan93,Caligari95} and the traditional $\alpha$-effect would be suppressed, the Babcock--Leighton mechanism was invoked to
take its place, with a similar-looking $\alpha$-coefficient
having a different interpretation \citep{Durney97}. Since the
Babcock--Leighton mechanism primarily takes place near the surface,
the $\alpha$-coefficient corresponding to it is usually assumed
to be confined near the solar surface.  \citet{CSD95} simply took $S = \alpha B$, with $\alpha$
concentrated at the surface as expected.  Even with such a
source function which did not include magnetic buoyancy
explicitly, they were able to get a periodic solution because
of the term
$$\frac{1}{r} \left[ \frac{\pa}{\pa r}
(r v_r B) + \frac{\pa}{\pa \theta}(v_{\theta} B) \right]$$
in (3), which implied that the toroidal field generated at
the tachocline was advected by the meridional circulation
to the surface where the Babcock--Leighton mechanism operated
on it. \citet{Kuker01} also followed this
approach.

Although it is possible to construct a Babcock--Leighton dynamo
model in this way without explicitly including magnetic buoyancy,
this is certainly not very physical or satisfactory. When the
toroidal field becomes sufficiently strong, its rise time due
to magnetic buoyancy is expected to be much shorter than the
advection time  by the meridional circulation.  So magnetic buoyancy
is expected to dominate over such advection and has to be included
in the model to make it more realistic.  In the next Section,
we discuss two popular procedures for incorporating magnetic
buoyancy---a non-local procedure and a local procedure. We first
discuss how these procedures are used in order to obtain regular dynamo
solutions.  Then we shall point out in \S~4 that we get into
added complications when we want to study irregularities of
the solar cycle.

\section{Non-local and local treatment of magnetic buoyancy}

The methods of specifying magnetic buoyancy in 2D kinematic
models of the flux transport dynamo can be broadly classified
into two categories: non-local and local.  Since the non-local
treatment of magnetic buoyancy is somewhat simpler, we discuss
that first.

\subsection{Non-local magnetic buoyancy}

The Babcock--Leighton mechanism essentially involves the
generation of the poloidal field near the surface from the
strong toroidal field at the bottom of the convection zone,
which has risen due to magnetic buoyancy. A simple way of
incorporating it is to take the source term $S(r, \theta,t)$
in (2) to have the form
$$S (r,\theta, t) = \alpha (r, \theta) B(r = r_{\rm bot}, \theta,t),
\eqno(4)$$
in which we usually take $\alpha (r, \theta)$ to be significantly
non-zero only near the surface, to ensure that $S(r, \theta,t)$
makes a contribution only near the surface.  Also, to get
$S(r, \theta,t)$ near the surface, we multiply $\alpha (r, \theta)$
not by the toroidal field $B$ there, but by the toroidal field
$B(r= r_{\rm bot}, \theta, t)$ at the bottom of the convection
zone. Since the rise time due to magnetic buoyancy is small
compared to the period of the dynamo, we normally use the same $t$  in $S$ and $B$ without introducing any time delay.

To the best of our knowledge, the method of treating magnetic
buoyancy in this way was first proposed by \citet{CD99} in their study of the evolution of the solar poloidal field. Afterwards, \citet{DC99} adopted it
for their flux transport dynamo model. Some other authors have
followed this procedure in their dynamo calculations since
that time \citep{CD2000,Guerrero04,CSZ05,Hotta10}.  

This method of treating magnetic buoyancy is a simple, robust
and stable method. It is found that dynamo models based on
this method of treating magnetic buoyancy remain stable on
changing the values of basic parameters over wide ranges.
However, in spite of its simplicity and attractiveness, this
method has the following unphysical features.

(1) We expect the toroidal field to be unstable to magnetic
buoyancy only after it has become sufficiently strong.  However,
in the non-local method of treating buoyancy, even a very weak
toroidal field at the bottom of the convection zone starts
contributing to the source term in (2).

(2) As a result of buoyant rise from the bottom of the convection
zone, the toroidal field at the bottom of the convection keeps
getting weaker.  The simple method of treating magnetic buoyancy
described above does not incorporate this effect and most of the authors
who treated magnetic buoyancy in this way did not allow
the weakening of the toroidal field due to magnetic buoyancy. 
As we shall see in the next Section, not doing this has serious
consequences when we study irregularities of the solar cycle
such as the Waldmeier effect.

\subsection{Local magnetic buoyancy}

In this approach, some toroidal magnetic field is transferred
from the bottom of the convection zone to the solar surface
and then this toroidal field at the surface is expected to
produce the poloidal field locally. Magnetic buoyancy is
known to be particularly destabilizing within the convection
zone \citep{Parker75,Moreno83}. So, as the toroidal
field evolves according to (3), we check at periodic intervals
if the toroidal field at any point within the convection zone
becomes stronger than a critical value $B_c$ and, if so, then
some toroidal field from such region is transferred to the
surface.  It may noted that this procedure introduces a limit
to the growth of the dynamo by not allowing $B$ to grow much
beyond $B_c$.  As a result, the dynamo can exhibit non-growing
oscillatory solutions even without introducing any kind of
quenching.  On the other hand, in the non-local procedure
described in \S~3.1, it is absolutely essential to include
some kind of quenching (a quenching of the $\alpha$-coefficient
being the most common) to stop the runaway growth of the magnetic
field. After the toroidal field is shifted to the top of the
convection zone, we have to prescribe some way of generating
the poloidal field from it.  Two ways of doing this are
described in the next two paragraphs.

{\em Local $\alpha$ parameterization.} One way is to prescribe the 
source term $S(r,\theta,t)$ in (2) simply 
as a product of $\alpha(r, \theta)$ confined around the surface and
the toroidal field $B (r, \theta, t)$ there, which has been shifted
there from those regions at a bottom of the convection zone where
 $B$ exceeded $B_c$ in a way that ensured the conservation of
the toroidal flux (i.e.\ the amount of toroidal flux deposited
near the surface has to equal the amount of toroidal flux
removed from the bottom of the convection zone). This method
has been followed in several publications from our group
\citep{Nandy01,Nandy02,CNC04,CNC05,CC06,CCJ07,
GoelChou09,KarakChou11,KarakChou12,KarakChou13,HKBC15}.

{\em {Durney's double ring method.}} The ideas of \citet{bab61}
and \citet{Leighton69} were followed more closely by \citet{Durney95,Durney97}.
Due to the action of the Coriolis force, the rising
flux tube gets tilted \citep{Dsilva93} and produces
two sunspots of opposite polarity at slightly different
latitudes.  In an axisymmetric 2D formulation, we have to
average over longitude, which gives us two flux rings of
opposite sign at two slightly different latitudes. The generation
of the poloidal field is prescribed in the following way.
Whenever $B$ in some region within the convection exceeds
a critical value $B_c$, we assume that a part of this $B$ gives
rise to the flux ring above the region and we put $A$ appropriate
for this flux ring in (2).  Here we do not discuss the details
of how we find $A$ appropriate for a flux ring, except to
mention that some assumption has to be made about the magnetic
field structure below and above the flux ring.  Presumably
the bipolar sunspot pair eventually gets disconnected from the toroidal
flux ring at the bottom of the convection zone from which it
formed, though it is not clear at the present time how and when
this disconnection takes place \citep{Longcope02}.
Figure~1 from \citet{Munoz10} shows the typical poloidal 
field structure assumed by them in their dynamo model with
Durney's double ring algorithm.

\begin{figure}
\begin{center}
\includegraphics[width=10 cm]{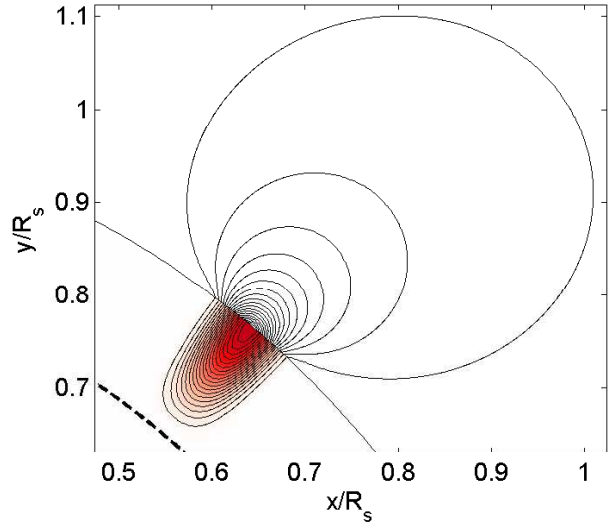}
\end{center} 
\caption{The poloidal field lines of a double ring formed
near the surface due to magnetic buoyancy.
Taken from \citet{Munoz10}.}
\end{figure}
It is found that the above two methods---local $\alpha$
parameterization and Durney's double ring method---give
qualitatively similar results \citep{Nandy01}.
However, \citet{Munoz10} have argued that the
double ring method is a superior method from the conceptual
point of view.

\subsection{Comparison between the non-local and local
methods}

It would have been wonderful if the non-local and local methods
of treating magnetic buoyancy gave qualitatively similar
results.  Unfortunately, that is not the case! When this was
first discovered by \citet{CNC05}, it
came as a surprise to most of the researchers in the field,
though now from hindsight we feel that this should have been
an expected result.  Figure~2 shows the poloidal field configurations
in two dynamo models in which magnetic buoyancy is treated by non-local
and local ($\alpha$ parameterization) methods, but which are 
identical dynamo models in all other respects.  When we use
non-local buoyancy, even the weak toroidal fields produced at the
high latitude start giving rise to the poloidal field. This
causes a multi-lobe structure of the poloidal field in this
case, shortening the period of the dynamo.  While the model
with the local $\alpha$ parameterization gives a period of
14 years, we get a period of barely 6.1 years on using the
non-local treatment of magnetic buoyancy.

\begin{figure}
\begin{center}
\includegraphics[width=10 cm]{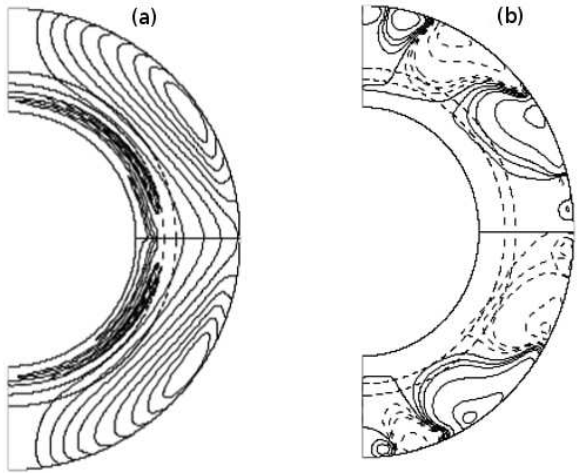} 
\end{center}
\caption{Poloidal field lines at a particular instant of
time in two solar dynamo simulations.  Both use the same  combination of parameters which were used to generate the
standard model in \S~4 of \citet{CNC04}.
The only difference between the two runs is that (a) was
generated by using local buoyancy treatment and (b) was generated
by using the non-local buoyancy treatment.  Whereas (a) is
taken from Fig.~14 of \citet{CNC04},
(b) is taken from Fig.~1 of \citet{CNC05}.}
\end{figure}

Because of these big differences, we cannot avoid the question
as to which of these two methods is better.  We first note that
both the methods are gross over-simplifications of a complicated
3D process.  In this sense, neither of these two methods can be
taken as a realistic depiction of magnetic buoyancy.  We believe
that the local treatment is a more physical and realistic depiction
of what actually happens in the Sun---a conclusion that will
be further reinforced from the discussion of the next Section.
We have used this method of local treatment
in most of the calculations done in our group.  Unfortunately,
this method is less robust and less stable than the non-local method
and seems to give results only within a narrow range of various
important dynamo parameters.  So, when we want to vary our
dynamo parameters over a wide range or when we use unusual
parameter specifications (such as a multi-cell meridional
circulation), we are often unable to obtain results with the
local $\alpha$ parameterization procedure and are forced to use
the more robust non-local procedure described in \S3.1 
\citep{KKC14,HKC14}.

\section{Modelling irregularities of the solar cycle}

The previous section outlined the various ways of treating
magnetic buoyancy and pointed out the differences in the
periodic dynamo solutions we get with them.  Within the last
few years, one of the goals of solar dynamo theory has been to
model the irregularities of the solar cycle. Only very recently
we realized that different formulations of magnetic buoyancy
may give radically different results in this important field
of study---especially when we consider irregularities of the
solar cycle caused by the variations in the meridional 
circulation.

The duration of the solar cycle becomes more when the meridional
circulation slows down \citep{DC99,Yeates08}.
So it is expected that variations
in the meridional circulation will introduce irregularities
in the cycle.  When this problem is studied with the help
of local $\alpha$ parameterization, the theoretical results
are broadly in agreement with observational data \citep{Karak10,KarakChou11,CK12,KarakChou13,HKBC15}. 
Before explaining how things change on changing the method of treating magnetic
buoyancy, let us say a few words about the basic physics of
the problem.  When a cycle becomes longer due to the slowing
of the meridional circulation, two competing effects take place.
The differential rotation has more time to generate more
toroidal field, trying to make the cycle stronger.  On the
other hand, diffusion has also more time to act on the magnetic
field and tries to make the cycle weaker. Which of these two
effects wins over depends on the assumed value of the turbulent
diffusion coefficient in the convection zone.

Most of the calculations in our group were done with a value
of turbulent diffusion around $10^{12}$ cm$^2$ s$^{-1}$ consistent
with mixing length argument (Parker, 1979, p. 629; Jiang et al. 2007). 
With such a value of turbulent diffusion,
the effect of diffusion trying to make the longer cycle weaker wins
over the effect of differential rotation trying to make longer
cycles stronger.  As a result, longer cycles tend to be weaker.
This naturally leads to an explanation of the Waldmeier effect
that the rise time of the cycle is inversely correlated with
the strength of the cycle \citep{KarakChou11}. 
On the other hand, several papers from HAO \citep{DC99,CD2000,DG06} used a value
of turbulent diffusion about 50 times smaller.  With such a
value of turbulent diffusion, the effect of differential rotation
trying to make longer cycles stronger wins over, giving the
opposite of the Waldmeier effect.

The better agreement with observational data on using the higher
value of turbulent diffusion clearly indicates that this higher
value must be closer to reality.  However, we realized only
recently that an appropriate handling of magnetic buoyancy is
also required to get a match with observational data.  We basically
need diffusion to be dominant so that longer cycles are weaker.
When we use a higher value of diffusion and treat magnetic 
buoyancy through local $\alpha$ parameterization, this happens
and we are able to explain effects like the Waldmeier effect
beautifully.  However, when we use the non-local buoyancy
method, we get into trouble even on taking the higher value of
diffusion.  In our calculations with local $\alpha$ parameterization,
toroidal flux is removed from the bottom of the convection zone
as a result of magnetic buoyancy.  On the other hand, in
the non-local buoyancy formulation used in the papers of the
HAO group \citep{DC99,CD2000}, the toroidal flux is never depleted as a result of
magnetic buoyancy.  Because of this, the toroidal field keeps
becoming stronger when the cycle is longer and, even with a
high value of diffusion, we find that longer cycles tend to
be stronger, giving the opposite of the Waldmeier.

\begin{figure}
\begin{center}
\includegraphics[width=10cm]{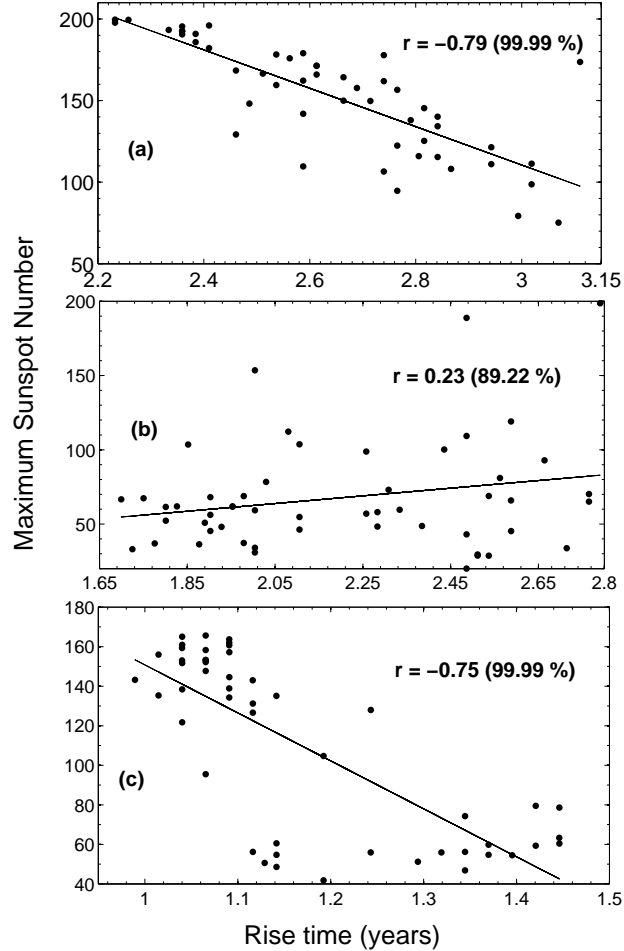} 
\end{center}
\caption{The correlation between the rise and the strength\
of the solar cycle from a theoretical solar dynamo model using
the same combination of parameters as the high diffusivity model
presented by \citet{KarakChou11}. Only fluctuations
in the meridional circulation are included.  The panels (a)
and (b) are obtained by using respectively the local treatment and
the non-local treatment (without depletion in the toroidal flux)
of magnetic buoyancy. The last panel (c) obtained by using
the non-local treatment of magnetic buoyancy in which the toroidal
field is not allowed to grow larger than a critical value $B_c$.
The panel (a) essentially the same plot
as Fig.~4(a) of \citet{KarakChou11} obtained with a different realization of the randomness.}
\end{figure}

Since this aspect of the problem was realized only recently
and has not yet been reported in regular journal publications,
we present some new results here.  Figure~3(a) shows the Waldmeier
effect (i.e.\ the anti-correlation between the rise time and
the strength of the cycles) obtained with a dynamo model with
a high value of diffusion in the convection zone on using the
local $\alpha$ parameterization.  If we treat magnetic buoyancy
through the non-local method as given by (4), while keeping
all other things of the dynamo unchanged, then we get the
opposite of the Waldmeier effect, i.e.\ a correlation rather
than an anti-correlation.  This is shown in Figure~3(b). To verify
that this is indeed caused by not including the depletion of
the toroidal field by magnetic buoyancy, if we now modify the
non-local method slightly by putting the restriction that the
toroidal field is not allowed to grow beyond a limiting value
$B_c$ (i.e.\ we put a specification in the code that whenever
$B > B_c$ we set $B = B_c$), then we again get back the Waldmeier
effect.  The result is shown in Figure ~3(c) which is obtained
by using the non-local buoyancy with the restriction that $B$ is
never allowed to grow beyond $B_c$.
We thus have a rather peculiar situation.  Although the
non-local magnetic buoyancy method is more robust and works
for a wide range of parameters, if this method is used blindly
while studying irregularities caused by the variations in the
meridional circulation, we are likely to get completely wrong
results.  This should make it clear that none of the presently
used methods of treating magnetic buoyancy is completely
satisfactory.  Each method has its limitations and we have
to keep these limitations in mind when interpreting the results
obtained with a particular method. 

\section{The latitudinal distribution problem}

So far we have discussed problems of a realistic 2D formulation
of magnetic buoyancy, which is an intrinsically 3D process. Now
we mention another problem connected possibly with magnetic
buoyancy, which is still very poorly understood. In a helioseismic
map of differential rotation (see, for example, \citet{Schou98}), 
it is clearly seen that differential rotation is concentrated 
more strongly at high latitudes than at low latitudes---the sign
being different at high and low latitudes.  Because of this, the
generation of the toroidal magnetic field from the poloidal magnetic
field is supposed to be more pronounced at the high latitudes.
At the low latitudes, the differential rotation present there first
has to `unwind' the toroidal field produced by the differential
rotation of the opposite sign at high latitudes and brought to the
low latitudes by the equatorward meridional circulation.  Only after
that, the differential rotation at low latitudes can build up the
toroidal field.  Hence, when we use the differential rotation
discovered by helioseismology, there is a propensity of stronger
toroidal field being produced at the high latitudes. If this
strong toroidal field is allowed to rise to the surface there
due to magnetic buoyancy, then we find sunspots at latitudes
higher than where they are seen.  The first authors to use
helioseismically determined differential rotation already noticed
this problem \citep{DC99, Kuker01}  

\citet{Nandy02} proposed a solution to this problem.
They suggested that the meridional circulation penetrates slightly
below the bottom of the convection zone, where the temperature
gradient is stable against convection and magnetic buoyancy
is suppressed to a large extent.  If this is the case, then
the strong toroidal field created in the high-latitude tachocline
may be pushed below the bottom of the convection zone and may not
be able to rise to the surface due to the suppression of magnetic
buoyancy there, thereby inhibiting formation of sunspots at high
latitudes.  Then the toroidal field would be advected equatorward
by the meridional circulation through layers slightly below the
bottom of the convection.  Then, when the penetrating meridional
circulation again enters the convection zone at low latitudes, the toroidal field
is brought into the convection zone and magnetic buoyancy again
takes over to produce sunspots at low latitudes.  This 
Nandy--Choudhuri hypothesis became a source of 
controversy---\citet{GM04} arguing that it is not
possible for the meridional circulation to penetrate into the
stable layers below the bottom of the convection zone, whereas
\citet{GB08} pointed out that such penetration is
possible. In many dynamo calculations from our group, we have
made the meridional circulation slightly penetrating in order
to confine sunspots to low latitudes. As \citet{CCC09} pointed out, one strong support for the
Nandy--Choudhuri hypothesis comes from the observation that
torsional oscillations begin at higher latitudes before the
start of a sunspot cycle.  Since torsional oscillations are
presumably driven by the Lorentz force of the dynamo-generated
magnetic field, this observation clearly suggests that the
strong toroidal field builds up in the high-latitude tachocline
a few years before this field is advected to low latitudes and
is able to produce sunspots, in accordance with the Nandy--Choudhuri
hypothesis.

In a study of the evolution of the poloidal field (in which
the dynamo equation was not solved), \citet{DC94}
mimicked the behaviour of the dynamo by restricting the source
term of the poloidal field to low latitudes.  Within the last
few years, several authors have solved the dynamo equation
by artificially restricting the Babcock--Leighton source term
to high latitudes in this fashion \citep{Hotta10, Munoz10}.
While this procedure produces nice-looking butterfly diagrams with
sunspots restricted to low latitudes, these authors provide
no physical justification for this procedure and we are not
sure whether this procedure is valid from the point of view
of basic physics.  In our opinion, we still do not have a
proper understanding of why sunspots are restricted only to
low latitudes.  Although this is an old problem and no significant
progress has been made in the last few years, we decided to
add this Section because the existence of this problem is
nowadays often not appreciated or acknowledged.  We should keep
this problem in mind and should not sweep it under the rug.

\section{Conclusion}

Although we so far do not have anything that can be called
the standard model of the solar dynamo \citep{Chou08}, 
flux transport dynamo models developed by different groups
have lots of common features. While there is a controversy
on the nature of the meridional circulation at present, it
is still not clear whether a serious revision of existing dynamo
models will be required. Another uncertainty about the value
of turbulent diffusion seems to be resolved now in favour of
higher diffusivity (such that the diffusion time scale is a few
years) because only with such diffusivity it is possible to
model various aspects of cycle irregularities \citep{Chou15}. 
Whether turbulent pumping is important for the solar
dynamo is another question. However, since the inclusion of
turbulent pumping does not change the results of high-diffusion
dynamo models too much \citep{KarakNandy12}, this is presumably
not a large source of uncertainty.  

As we have discussed here, the biggest uncertainty in the flux
transport dynamo at the present time arises from the treatment
of magnetic buoyancy. There have been two widely used procedures
for treating magnetic buoyancy in 2D kinematic models of the flux
transport dynamo: (i) the non-local treatment
in which the toroidal field at the bottom of the convection zone is
assumed to contribute directly to the generation of the poloidal field,
without itself being depleted, and (ii) the local treatment in which 
a part of the toroidal
field from the bottom of the convection zone is shifted to the top
whenever it exceeds a critical value. With magnetic buoyancy treated in either
of these ways, it is possible to arrange the parameters of the model
in such a way that the dynamo solution reproduces various characteristics
of the solar cycle.  However, for the same set of parameters, we get
very different results on using these different treatments of magnetic
buoyancy.  The non-local treatment is more robust and is preferred when
we want to study the behaviour of our system over a wide range of
parameters.  On the other hand, the local treatment with the toroidal
field depletion at the bottom of the convection zone allows the proper
reproduction of results under irregular situations (such as the Waldmeier
effect) and appears the more physically realistic treatment.

This is certainly an unsatisfactory situation and presumably we cannot
model magnetic buoyancy sufficiently well if we restrict ourselves
only to strictly 2D situations.  It may be noted that, in the early
studies of magnetic buoyancy based on thin flux tube equation \citep{Spruit81, Chou90}, 
some physical effects could be studied
through 2D calculations \citep{CG87, CD90, DC91}.
However, a proper study of the tilt of bipolar sunspot regions (which is responsible for
the Babcock--Leighton process) could be carried out only when
the calculations were carried out in 3D \citep{Dsilva93, Fan93, Caligari95}.
Essentially, magnetic buoyancy is inherently a 3D process and
cannot be included very satisfactorily in 2D kinematic dynamo
models.  Further advances in 2D kinematic models are unlikely to
solve the problems we are facing now.

The ultimate goal of solar dynamo models is to construct fully
dynamically consistent 3D models, in which the velocity fields
are also calculated from the fundamental equations \citep{Karakreview14}.  
Beyond 2D kinematic models, however, there is an 
intermediate possibility---3D kinematic models, in which the
mean velocity fields (such as differential rotation and meridional
circulation) are supposed to be axisymmetric and are specified
(i.e.\ not calculated from the basic fluid equations), but the
magnetic field is treated in a full 3D fashion and is allowed to
evolve non-axisymmetrically so that we are able to model the
formation of tilted bipolar sunspots and the Babcock--Leighton
process more realistically.  The relatively few citations (according
to ADS) to the first paper suggesting this approach \citep{YM13} 
indicate that the importance of this approach is not generally
recognized.  But we believe that this is going to be the next
important step in flux transport dynamo theory beyond 2D kinematic
model.  Only a few authors have so far presented results with this approach \citep{MD14}.
While this approach, when developed and integrated with a 
realistic dynamo model, should be more satisfactory than the currently widely used two 
treatments of magnetic buoyancy in 2D kinematic models. However,
a 3D kinematic model is not expected to immediately 
solve all the problems connected with magnetic buoyancy 
in the 2D kinematic model. For example, we have
discussed in \S~5 the problem of why sunspots appear in lower
latitudes.  This problem will not be solved with the 3D formulation
of magnetic buoyancy.  Rather, we shall need a better understanding
of the physics of the tachocline to address this question.  We
thus expect a combination of more realistic calculation of 
magnetic buoyancy coupled with a better understanding of the
tachocline to put the flux transport dynamo on a more satisfactory
footing.

\section*{Acknowledgements}
We thank Bidya Binay Karak for valuable discussions. 
This work is partly supported by DST through the J.C. Bose Fellowship awarded to ARC.
GH thanks organizers of NASA LWS Workshop On Solar Dynamo Frontiers and Flux Emergence Workshop
for providing travel support. GH also thanks CSIR india for financial support.

\section*{References}
\bibliography{myref}

\end{document}